\documentclass[11pt]{article}
\usepackage{a4wide}

\usepackage[dvipdfmx,hiresbb]{graphicx} 
\usepackage{mediabb}                    

\usepackage{amsmath,amssymb}
\usepackage{color}
\usepackage[bookmarks,bookmarksnumbered]{hyperref}
\usepackage{cellspace}
\begin{document}
\title{\vbox{
\baselineskip 14pt 
\hfill 
} \vskip 1cm
\bf \Large 8d Supergravity, Reconstruction of Internal Geometry and the Swampland
\vskip 0.5cm
}
\author{
{\bf Yuta Hamada}\thanks{E-mail: \tt yhamada@fas.harvard.edu}\,  and {\bf Cumrun Vafa}\thanks{E-mail: \tt vafa@g.harvard.edu}
\bigskip\\
\it\normalsize
Jefferson Physical Laboratory, Harvard University, Cambridge, MA 02138, USA\\
}
\date{}

\maketitle  
\vspace*{1cm}  
\begin{abstract} 
We sharpen Swampland constraints on 8d supergravity theories by studying consistency conditions on worldvolume theory of 3-brane probes.
Combined with a stronger form of the cobordism conjecture, this leads to the reconstruction of the compact internal geometry and implies strong restrictions on the gauge algebra and on some higher derivative terms (related to the level of the current algebra on the 1-brane).
In particular we argue that 8d supergravity theories with $\mathfrak{g}_2$ gauge symmetry are in the Swampland.
These results provide further evidence for the string lamppost principle in 8d with 16 supercharges.
\end{abstract} 

\setcounter{page}{1} 
\newpage

\section{Introduction}
\label{Sec:intro}

The goal of the Swampland program~\cite{Vafa:2005ui} (see \cite{Brennan:2017rbf,Palti:2019pca,vanBeest:2021lhn} for reviews) is to find constraints on field theories which can couple consistently to quantum gravity.
One of the tools to elucidate Swampland bounds is to study consistency conditions on worldvolume theory on brane probes which was initiated in \cite{Kim:2019vuc} for the 1-brane probes.
Since the two-form bulk field exists for supersymmetric theories, the completeness hypothesis requires an object coupled to it, 1-brane. 
Using the anomaly inflow to compute the central charge of 1-brane probes and the assumption of supersymmetry on the probes, one can derive the constraints on the gauge algebra for theories with 16 supercharges~\cite{Kim:2019vuc,Kim:2019ths}. 
Similarly, 5 and 6d theories with 8 supercharges are constrained~\cite{Kim:2019vuc,Lee:2019skh,Katz:2020ewz}. 
These developments reveal that, at least for 16 supercharges, only a finite set of field theory matter content can be consistently coupled to quantum gravity since the rank of the possible gauge group is bounded from above. 

However, the rank realized in string theory is more restricted. 
With 16 supercharges, the rank of string compactifications to nine dimensions is only $1, 9$ or $17$, and to eight dimensions is only $2, 10$ or $18$.
This pattern is explained in \cite{Montero:2020icj} by using the anomaly cancellation and the cobordism conjecture~\cite{McNamara:2019rup}. 
The cobordism conjecture is one of the Swampland criteria which states that all the cobordism classes in a consistent theory of quantum gravity vanish. 
This is understood as a generalization of the completeness hypothesis and absence of global symmetry because violation of one of them leads to the non-trivial cobordism classes.

These results support the String Lamppost Principle (SLP), which states that all consistent theories of quantum gravity are in the string landscape. 
On the other hand, the Swampland constraints so far are mainly constraints on the rank of the gauge group.
In order to establish SLP, a natural next step is to give bounds on the actual gauge group, rather than the rank.  See \cite{Cvetic:2020kuw,Montero:2020icj,Dierigl:2020lai} for some progress in this direction.
However many questions remain open.
For example, in string compactifications to eight dimensions, the gauge algebra $18\,\mathfrak{su}(2)$ does not show up while $6\,\mathfrak{su}(4)$ does appear although these two have the same rank $18$. 
Similarly, there are no known string theory constructions of 8d supergravity theories with $\mathfrak{g}_2$ gauge symmetry although there are no known Swampland arguments to exclude them.
Can we explain these facts without relying on string theory? 

In this paper, we provide positive answers to these questions.  
We obtain Swampland constraints by requiring the consistency of 3-brane probe theory on the magnetic brane associated to the 2-form gauge field of supergravity theories in 8 dimensions.  The completeness hypothesis requires the existence of the 3-brane probe.
In the presence of non-abelian gauge fields in 8 dimensions, the bulk theory admits gauge instanton solutions, and these also carry the same 3-brane charge.
To place bounds on the gauge groups, we focus on the instanton 3-brane in the 8d supergravity with 16 supercharges (see~\cite{Sezgin:1982gi,Awada:1985ag}
for a discussion of these 3-branes).

When there are multiple non-abelian gauge factors, each of the non-abelian factors has its own gauge instanton configuration. 
A stronger form of the cobordism conjecture~\cite{Cobordism}, 
applicable to theories with higher supersymmetries, implies that the instanton 3-branes must be connected through {\it supersymmetric} deformations which we argue for the case at hand is a one-dimensional Coulomb branch. 
We then derive the bounds on gauge algebras and related couplings by requiring consistency conditions on this Coulomb branch.
These allow for the existence of $6\,\mathfrak{su}(4)$ but exclude $18\,\mathfrak{su}(2)$, and therefore provides further evidence for the SLP.
Moreover, we argue that theories with $\mathfrak{g}_2$ gauge symmetry are in the Swampland.

The restrictions we derive have a clear interpretation in F-theory on elliptic $K3$~\cite{Vafa:1996xn}. 
Indeed, we show that the Coulomb branch of the instanton 3-brane probe is viewed as base $\mathbb{P}^1$ of the F-theory compactification with the elliptic fibration denoting the gauge coupling along the Coulomb branch.\footnote{This is the same as what happens in the D3-brane probing the F-theory \cite{Banks:1996nj}.}
In this sense, even though we start from 8d supergravity, we can reconstruct the compactified internal geometry, and obtain the bound inherited from the higher dimensional theory!
We believe that the reconstruction of the internal geometry and in particular its compactness (which we argue should be generally true for an arbitrary brane probe) can be investigated more broadly, and is a promising new direction for the Swampland program.

The organization of this paper is as follows. In Section~\ref{Sec:Review} we review known string constructions and Swampland bounds for 8d supergravity theories.
In Section~\ref{Sec:bound} we first discuss properties of instanton 3-brane in 8d, which play a central role in our argument and then derive bounds on the gauge algebras.
In Section~\ref{Sec:power} we show the power of the restrictions we have found.
In Section~\ref{Sec:conclusions} we present our conclusion.

\section{String constructions and Swampland constraints of 8d theories with 16 supercharges}
\label{Sec:Review}
In this section, we review string constructions and Swampland bounds of 8d theories with 16 supercharges.

\subsection{String constructions}
\label{Sec:String_construction}
Known 8d string theory compactifications with 16 supercharges have gauge groups whose rank is 18, 10, or 2. 
Here we briefly review these constructions. See also \cite{Dabholkar:1996pc,deBoer:2001wca,Taylor:2011wt,Kim:2019ths,Montero:2020icj} for string constructions of 8d theories.

\begin{itemize}
\item Rank $=18$

Theories with Rank $=18$ are realized by F-theory on elliptic $K3$~\cite{Vafa:1996xn}, and type I or heterotic string on $T^2$.
The full list of the type of singularity of the elliptic $K3$, which leads to the allowed gauge algebras is found in \cite{2005math......5140S}.
Similarly for the heterotic string on $T^2$, all the groups of maximal enhancement are listed in \cite{Font:2020rsk} which agrees with \cite{2005math......5140S}. 

\item Rank $=10$

Theories with Rank $=10$ are realized by F-theory on elliptic $K3$ with one frozen singularity ($O7_+$ plane)~\cite{Witten:1997bs}, \footnote{We use the notation where $O7_+$ has the positive $D7$ charge.}  CHL string~\cite{Chaudhuri:1995fk,Chaudhuri:1995bf,Mikhailov:1998si}, IIA orientifold on Mobius strip, and type I or heterotic string on $T^2$ without the vector structure~\cite{Witten:1997bs}.
The rank $10$ gauge algebras without $\mathfrak{u}(1)$ are listed in Tables~\ref{Tab:Rank10} and \ref{Tab:Rank10_2} in Section~\ref{Sec:power}. For eight dimensional CHL string, all groups of maximal enhancement are listed in \cite{Font:2021uyw}. Our tables agree with \cite{Font:2021uyw}.\footnote{In the arXiv version one, we missed $\mathfrak{e}_7\oplus\mathfrak{sp}(1)\oplus\mathfrak{su}(3)$, as pointed out in \cite{Font:2021uyw}. In our case, this comes from No.2994 in \cite{2005math......5140S}.} 

\item Rank $=2$

Theories with Rank $=2$ are realized, for example via F-theory on elliptic $K3$ with two $O7_+$ planes~\cite{Witten:1997bs}.
Other realizations are M-theory on $(\text{Klein bottle})\times S^1$ \cite{Aharony:2007du} and Dabholkar-Park background~\cite{Dabholkar:1996pc}.
The gauge algebras are listed in Table~\ref{Tab:Rank2} in Section~\ref{Sec:power}.

\end{itemize}

Note that in all these three cases the F-theory geometry involves an elliptic $K3$.  We will be able to explain this fact directly by the study of 3-brane moduli space from the perspective of the eight dimensional effective theory.

\subsection{Review of the Swampland bounds}
\label{Sec:swampland_review}
In 8d, the perturbative gauge anomaly is computed from the 1-loop pentagon diagrams. 
However, this anomaly vanishes for supergravity since only chiral fields with adjoint representation appears~\cite{Taylor:2011wt}. 
At this level, any gauge symmetry is consistent. Nevertheless, there is a number of Swampland constraints, as we summarize below.

\begin{itemize}
\item The rank must be even because of a global gravitational anomaly~\cite{AlvarezGaume:1983ig,Witten:1985xe}.

\item The gauge algebras $\mathfrak{so}(2n+1),\, \mathfrak{f}_4$ suffer from a global gauge anomaly \cite{Garcia-Etxebarria:2017crf}. 
The $\mathfrak{sp}(n)$ algebra has a subtle anomaly,\footnote{We use the notation $\mathfrak{sp}(1)=\mathfrak{su}(2)$.} which can be canceled by coupling it to a TQFT. To cancel the anomaly, TQFT gives the constraint that allowed $\mathfrak{sp}(n)$ instanton numbers are even~\cite{Pantev:2005rh,Pantev:2005wj,Pantev:2005zs,Hellerman:2006zs,Seiberg:2010qd,Tanizaki:2019rbk}.

\item Based on $1$-brane unitarity, the rank of the gauge group is bounded from above~\cite{Kim:2019ths}.
The bulk gauge algebra $\mathfrak{g}=\sum_i \mathfrak{g}_i$ corresponds to the level $\ell_i$ current algebra in 1-brane probe. The central charge of the current algebra is bounded as
\begin{align}\label{Eq:1-brane}
c_\mathfrak{g}=\Sigma_i \frac{\ell_i \, d_{\mathfrak{g}_i}}{\ell_i+h^\vee}\leq
\begin{cases}
18 \quad \text{for $\kappa=1$}\\
2 \quad \text{for $\kappa=0$}
\end{cases},
\end{align}
where $d_\mathfrak{g}$ is the dimension, $h^\vee$ is the dual Coxeter number, and $\kappa=0$ or $1$. 
See Table~\ref{Tab:Group} for concrete numbers of $c_\mathfrak{g}$.
The positive integers $\ell_i$ and $\kappa$ appear in the Bianchi identity of the 3-form flux $H_3$:
\begin{align}\label{Eq:Bianchi}
dH_3 = \ell_i \mathrm{Tr}F_i^2 +\kappa\,\mathrm{tr}R^2.
\end{align}
Combined with the anomaly inflow, the bound \eqref{Eq:1-brane} leads to 
\begin{align}
\text{Rank}\leq
\begin{cases}
18 \quad \text{for $\kappa=1$}\\
2 \quad \text{for $\kappa=0$}
\end{cases}.
\end{align}
\begin{table}[!ht]
\begin{center}
\small
\begin{tabular}{|c||cccccccc|}\hline
 $\mathfrak{g}$ &  $\mathfrak{su}(n)$  & $\mathfrak{so}(2n)$ & $\mathfrak{sp}(n)$ & $\mathfrak{g}_2$  & $\mathfrak{f}_4$ & $\mathfrak{e}_6$ &$\mathfrak{e}_7$ & $\mathfrak{e}_8$\\\hline\hline
$r$  & $n-1$ & $n$ & $n$ & $2$ & $4$ & $6$ & $7$ & $8$ \\\hline
$h^\vee$  & $n$ & $2n-2$ & $n+1$ & $4$ & $9$ & $12$ & $18$ & $30$ \\\hline
$d_\mathfrak{g}$   &  $n^2-1$ & $n(2n-1)$ & $n(2n+1)$ & $14$ & $52$ & $78$ & $133$ & $248$ \\\hline
$\frac{d_\mathfrak{g}}{18(1+h^\vee)}$  & $(n-1)/18$ & $n/18$ & $\frac{n}{9}\left(1-\frac{3}{2(n+2)}\right)$ & $7/45$ & $13/45$ & $1/3$ & $7/18$ & $4/9$ \\\hline
$\frac{d_\mathfrak{g}}{9(2+h^\vee)}$  & $\frac{1}{9}\left(n-1-\frac{n-2}{n+2}\right)$ & $(2n-1)/18$ & $\frac{n}{9}\left(2-\frac{5}{n+3}\right)$ & $7/27$ & $52/99$ & $13/21$ & $133/180$ & $31/36$ \\\hline
$\frac{\text{order}(\Delta)}{24}$  & $n/24$ & $(n+2)/24$ & $(10+n)/24$ & - & - & $1/3$ & $3/8$ & $5/12$ \\\hline
\end{tabular}
\caption{
Group theory factor. Here $r, h^\vee$, and $d_\mathfrak{g}$ are the rank, dual Coxeter number, and the dimension of the group, respectively. 
The third and fourth rows are the central charges of the current algebra \eqref{Eq:1-brane} divided by $18$ (which is the maximum allowed by unitarity) with levels one and two, respectively.
The last row is related to a 3-brane bound we will provide, see Section~\ref{Sec:bound} and discussion around \eqref{Eq:result}\eqref{Eq:result_2}.
}
\label{Tab:Group}
\end{center}
\end{table}

\item Based on the cobordism conjecture~\cite{McNamara:2019rup} together with anomaly cancellation, consistent theories must have~\cite{Montero:2020icj} 
\begin{align}
\text{Rank}=2,\,10,\,18.
\end{align}

\item Constraints on the global structure of the gauge group are provided in \cite{Montero:2020icj,Cvetic:2020kuw}. 
There are non-trivial constraints on the non-simply connected groups. 
See Table 2 of \cite{Montero:2020icj} and equation (6) in \cite{Cvetic:2020kuw} for the concrete bounds.

\item An open question is whether 8d supergravity with $\mathfrak{g}_2$ gauge symmetry is in the landscape or the Swampland.

\end{itemize}

To summarize, the gauge algebras allowed by the Swampland bounds so far are the following.
\begin{enumerate}
\item Rank $=18$ with $\kappa=1$

Only the simply-laced algebras with the level one or $\mathfrak{u}(1)$'s are allowed. 

\item Rank $=10$ with $\kappa=1$

In addition to the simply-laced algebras and $\mathfrak{u}(1)$, $\mathfrak{sp}(n)$ and $\mathfrak{g}_2$ are allowed.
The level is not constrained as long as \eqref{Eq:1-brane} is satisfied.
There are no known string theory constructions with $\mathfrak{g}_2$ symmetry.

\item Rank $=2$ with $\kappa=0$

Only the simply-laced algebras with the level one or $\mathfrak{u}(1)$'s are allowed.

\item Rank $=2$ with $\kappa=1$

There are no known string constructions for this case, but their existence has not been ruled out.

\end{enumerate}

\section{Derivation of the bound on the gauge algebra}
\label{Sec:bound}

In this section, we derive bounds on the gauge algebra by requiring the consistency of the Coulomb branch geometry of the 3-brane probe theory.
We first provide a general argument that the moduli space of brane probes should lead to a compact manifold (which we anticipate to have many more applications beyond the current project), and then make use of the exact results of 4d $\mathcal{N}=2$ theory (for which there are powerful tools available beginning with the work ~\cite{Seiberg:1994aj,Seiberg:1994rs}) to restrict the 3-brane probe Coulomb branch moduli.  In particular we show that the 3-brane probe geometry leads to an elliptic $K3$ geometry, which is the key to the bounds we will find.

\subsection{Moduli of brane probes and compact internal geometry}

Here we argue that the moduli space of brane probes is a compact space with a discrete spectrum for the Laplacian.  We will argue this for arbitrary $p$-branes for $p<d-2$.  We will present the argument in the context of supersymmetric theories, but we believe the result is more general. 

Let us first consider 0-brane probes.  These can be viewed as black holes.  Let ${\cal M}$ denote this moduli space. To find the spectrum of the black holes we need to find the energy spectrum of the sigma model on ${\cal M}$.  This is given by the spectrum of the Laplacian on ${\cal M}$.  Each eigenstate corresponds to a state of a black hole. Since the number of black hole states in any given mass range is finite (given by $\int dM\   {\rm exp}(S(M))$ where $S(M)$ is the BH entropy of mass $M$) we conclude that the spectrum of the Laplacian of ${\cal M}$ is discrete.  This in particular implies that ${\cal M}$ cannot be non-compact.
More generally consider the moduli of $p$-branes. We first compactify the supersymmetric theory on $p$-dimensional periodic torus, and consider the $p$-branes wrapped around the torus.  The moduli space of the resulting 0-brane will include as part of its moduli space the moduli of the $p$-brane and in addition the moduli which may arise by the degrees of freedom arising from its wrapped modes on the torus.  The discreteness of the spectrum of 
Laplacian for the resulting 0-brane moduli which is required by the finiteness of the black hole entropy implies that the spectrum of Laplacian is discrete for the $p$-brane moduli.  In particular the $p$-brane moduli is compact.

What we will see in the following is that for the case of the 3-branes of interest to us the compact moduli space is nothing but the internal geometry of string compactification.  In this case the black hole entropy being finite gets related to finiteness of $M_{pl}$ which is related to finiteness of the volume of the internal space. This idea also points to the fact that studying brane probe moduli in an effective low energy theory, can potentially lead to reconstructing the internal geometry of the string compactification which is somewhat surprising from the perspective of the low energy theory (as for example the spectrum of KK modes which are massive can be read off from such moduli space).

It is natural to expect that this argument can be extended to scalar fields in the bulk and to lead to effective ``compactness'' of scalar moduli fields,  leading to an argument for the distance conjecture~\cite{Ooguri:2006in} based on black hole entropy~\cite{Distance}.

\subsection{3-brane}
\label{Sec:3-brane}
8d supergravity theories come along with a 2-form gauge field in the supergravity multiplet which is dual to a four form gauge field which we denote by $B_4$.
In other words, magnetically charged brane under the 2-form gauge field, is a 3-brane.  By completeness hypothesis such a 3-brane should exist.
Moreover we make the stronger assumption, as in the 1-brane case, that this brane is BPS leading to ${\cal N}=2$ supersymmetry on its worldvolume, which is motivated by the fact that such 3-black brane solutions with ${\cal N}=2$ supersymmetry do exist in supergravity setup.
We would like to explore the moduli space of the worldvolume theory for such a 3-brane and connect it to possible gauge algebras.
The coupling of the gauge field to $B_4$ is  through a coupling
\begin{align}
 \int B_4\wedge \left(\ell_i\mathrm{Tr}F_i^2- \kappa\, \mathrm{tr}R^2\right), 
\end{align}
where $\ell_i$ and $\kappa$ are positive integers in \eqref{Eq:Bianchi}, and we normalize $\mathrm{Tr}F^2$ in such a way that the integral of the BPS instanton gives one, as in \cite{Kim:2019vuc}. 
The integer $\ell_i$ is identified as the level of the current algebra realized on the 1-brane probe~\cite{Kim:2019ths}, which is magnetically charged under $B_4$.
The gauge instanton 3-brane solution of the bulk has $B_4$ charge $\ell_i$. The fact that gauge instantons for each gauge group lead to the same universal 3-brane charge is one key fact that we will use.  Zero size instantons of a gauge group $G$ inherits $G$ as a global symmetry (see e.g. the discussion of \cite{Callan:1991dj,Callan:1991ky}).
The finite size instantons break this global symmetry.
Therefore, finite size instanton is described by the Higgs branch of the theory.  Note that this is a local fact about gauge instantons and is consistent with what one sees in string theory (with or without dynamical gravity) (see e.g. ~\cite{Witten:1995gx,Ganor:1996mu,Seiberg:1996vs}).

Next, we discuss the cobordism conjecture~\cite{McNamara:2019rup}: all the cobordism classes in a consistent theory of quantum gravity vanish. This is understood as a generalization of the completeness hypothesis and absence of global symmetry because violation of one of them leads to the non-trivial cobordism classes.
We use a stronger version of it: all the configurations having the same gauge charge with at least 8 supercharges are connected by a supersymmetric deformation\cite{Cobordism}.\footnote{Note that with lower number of supersymmetries they are also connected according to cobordism conjecture, but to get from one to the other we may have to go through configurations which break supersymmetry. For example one can generically go from one critical point of a superpotential (which is allowed for supersymmetric theories with say $4$ supercharges) to another, by passing through non-supersymmetric configurations.}

We use this stronger version of the cobordism conjecture in the following way. 
Suppose that there are two or more non-abelian gauge algebras. 
We can consider the instanton solution of each non-abelian factor. 
If these instantons have the same level ($B_4$ charge) $\ell_i=\ell$, the stronger cobordism conjecture states that these are connected by the moduli space.
This cannot be the Higgs branch.  To see this consider two gauge groups $G_1$ and $G_2$ and recall that the 3-brane can be realized by zero size instanton of either $G_1$ or $G_2$.  Consider a configuration of $G_2$ zero size instanton, and suppose we wish to get to it from the zero size instanton of $G_1$.  Since at the end there is no gauge configuration of $G_1$ turned on, the global symmetry of $G_1$ is not broken on the probe.  Since fattening the $G_1$ instantons is the Higgs branch, we thus see that we need to have gone on the Coulomb branch of the probe ${\cal N}=2$ theory.  Thus the 3-brane instanton of $G_1$ and $G_2$ are connected via a Coulomb branch of the 3-brane ${\cal N}=2$ supersymmetric probe theory.  Note that $G_i$ does not have to be non-abelian, as we can also consider instantons of abelian theories.  These will not have a  Higgs branch
but will also be connected via a Coulomb branch as argued above.

The next question is the dimension of the Coulomb branch. 
We expect that given the assumption of the Higgs branch and flavor symmetry, the structure of the Coulomb branch is highly constrained. 
Indeed, for SCFT case, the central charges are determined from this requirement~\cite{Beem:2013sza,Shimizu:2017kzs},\footnote{See also \cite{Beem:2019tfp,Beem:2019snk} for constraints on Coulomb branch from the known Higgs branch by using vertex algebra.} see Table~\ref{Tab:Higgs_Branch}.  Moreover, for a single instanton number through various local constructions such as ADHM and exceptional groups (and in particular Minahan-Nemeschasky theories~\cite{Minahan:1996fg,Minahan:1996cj} which are independent of having a dynamical quantum gravity) one can conclude that the dimension of Coulomb branch for the probe theory for all the simply-laced groups is 1. The case of $\mathfrak{sp}(n)$ and $\mathfrak{g}_2$ will be discussed later.

\begin{table}[!ht]
\begin{center}
\small
\begin{tabular}{|c||ccccccccc|}\hline
 $\mathfrak{g}$ &  $\mathfrak{su}(2)$ & $\mathfrak{su}(3)$  & $\mathfrak{so}(8)$ & $\mathfrak{sp}(n)$ & $\mathfrak{g}_2$  & $\mathfrak{f}_4$ & $\mathfrak{e}_6$ &$\mathfrak{e}_7$ & $\mathfrak{e}_8$\\\hline\hline
$a$  & $11/24$ & $7/12$ & $23/24$ & $n/24$ & $17/24$ & $4/3$ & $41/24$ & $59/24$ & $95/24$ \\\hline
$c$  & $1/2$ & $2/3$ & $7/6$ & $n/12$ & $5/6$ & $5/3$ & $13/6$ & $19/6$ & $31/6$\\\hline
$D(u)$  & $4/3$ & $3/2$ & $2$ & - & $5/3$ & $5/2$ & $3$ & $4$ & $6$\\\hline
\end{tabular}
\caption{
The central charges of SCFTs with flavor symmetry $\mathfrak{g}$ and the one-instanton moduli space as the Higgs branch. 
This table is taken from \cite{Beem:2013sza,Shimizu:2017kzs}. 
The last row $D(u)$ is computed by using $4(2a-c)=2D(u)-1$ for the rank $1$ theory~\cite{Shapere:2008zf,Argyres:2016xmc} (This formula is modified when the gauging of the discrete symmetry is involved~\cite{Argyres:2016yzz}). 
A theory with $\mathfrak{sp}(n)$ flavor symmetry is given by just a free half-hypermultiplet, see also Section~\ref{Sec:sp(n)}. 
Note that there are two possibilities of the central charges for theories with $\mathfrak{su}(2)=\mathfrak{sp}(1)$ flavor symmetry. 
One is the interacting SCFT ($\mathfrak{su}(2)$ in Table), another is the free SCFT ($\mathfrak{sp}(1)$ in Table).
}
\label{Tab:Higgs_Branch}
\end{center}
\end{table}

So far, we have argued that the instanton 3-brane probe worldvolume theory has the one-dimensional Coulomb branch.  The 3-brane charge of the instanton 3-brane of gauge group $\mathfrak{g}_i$ is given by the product of the level $l_i$ and the instanton number $n_i$:  
$$Q=l_in_i$$
The strong cobordism conjecture implies that the minimal 3-brane instantons have the same charge $Q$ and thus the minimal product $l_in_i$ should be the same for all the gauge groups.  For all the simply laced groups the minimum instanton number is $n_i=1$ which implies that the levels $l_i$ are all the same:
$$l_i=l_{ADE}$$
In the rank 18 case and the rank 2 case (with $\kappa=0$), the unitarity condition on the 1-brane probe leads to $l_i=1$.  In the rank 10 case higher levels can also occur.  As we will discuss later for the $\mathfrak{sp}(n)$ case (which occurs only for rank 10 case) only even instanton numbers are allowed and in this case we learn that
$2l_{\mathfrak{sp}(n)} =l_{ADE}$.

\subsection{The geometry of instanton 3-brane moduli}

Having found out that the 3-brane moduli is the Coulomb branch of $\mathcal{N}=2$ supersymmetric theory and also knowing that it has to be a compact space with a discrete spectrum for the Laplacian puts severe restrictions on it.  Here we first argue that this implies that the moduli space is a sphere and then we use this to obtain bounds on the type of the singularities it leads to, which in turn gets related to the gauge algebras that can appear.

The Coulomb branch geometry is special K\"ahler because of the $\mathcal{N}=2$ supersymmetry. 
At the generic point, the low energy theory is pure $U(1)$ gauge theory,\footnote{Except for the center of mass modes.} where the parameter of the theory is the complexified $U(1)$ gauge coupling $\tau$. 
This is well defined up to the electromagnetic duality transformation
\begin{align}
&\tau \to \frac{a \tau+b}{c \tau + d},
&&\begin{pmatrix}
a & b \\
c & d
\end{pmatrix}
\in SL(2,\mathbb{Z}).
\end{align}
The metric of the Coulomb branch is
\begin{align}
ds^2=\tau_2 \,dud\bar{u},
\end{align}
where $u$ is the local coordinate. 

As we have argued the Coulomb branch space parameterized by $u$ should be a compact 1-dimensional space.  This in particular implies that
\begin{align}
\int \mathcal{R} =2\pi(2-2g)
\end{align}
where $\mathcal{R}$ is the curvature of the $u$-space and $g$ is the genus of the compact space.  Due to special geometry we know that the metric on $u$ space is special and in particular $\mathcal{R}=\partial \overline{\partial} \log \tau_2$.
This leads to
\begin{align}
\label{Eq:inequality}
\int \mathcal{R} ={1\over 2}\int \frac{d^2\tau}{\tau_2
^2}  =2\pi( 2-2g)
\end{align}
Since the left-hand side is positive, we conclude that $g=0,1$ are the only possibilities.  We will now argue that the case with $g=1$ is not allowed (as long as we have any non-abelian gauge symmetry).  Indeed the associated Coulomb branch moduli near the zero size instantons has been studied and they lead to $u$-dependent $\tau$.  However,
if $g=1$ then $\tau$ would be constant.   So as long as we have any non-abelian gauge symmetry this is ruled out.\footnote{This is consistent with the fact that in the case of theories with 32 supercharges, the 3-brane moduli will have ${\cal N}=4$ supersymmetry and thus the moduli space would be flat, (and by compactness argument of brane probe a flat torus) which would indeed not be consistent with an instanton moduli space.  This is consistent with the fact that theories with 32 supercharges do not have matter gauge fields and thus no associated instantons.}.  We thus conclude that as long as we have any non-abelian gauge symmetry $g=0$.  The above integral should thus be $4\pi$ which means that the map from $u$-space to $\tau$ space must cover the fundamental domain 24 times (using the fact that the volume of upper half-plane moduli space is given by $\int {d^2\tau\over \tau_2^2}={4\pi\over 12}$).  This implies that if we consider the $U(1)$ gauge coupling viewed as the moduli $\tau$ of the elliptic fibration over the Coulomb branch, leads to a complex 2-dimensional space which is topologically the K3 manifold~\cite{Greene:1989ya}. The fibration is described by the Weiessstrass form,
\begin{align}
y^2 = x^3 + f(u) x + g(u),
\end{align}
where $f$ and $g$ are degree 8 and 12 polynomials, respectively.
In general, singularities appear at zeros of the discriminant $\Delta$:
\begin{align}
\Delta = 4f^3 + 27g^2.
\end{align}
Note that $\tau$ is not a globally well-defined function, but undergoes monodromy as we go over the Coulomb branch parameter $u$.  This is the F-theory geometry by identifying $\tau$ as the axio-dilaton~\cite{Vafa:1996xn}.  In other words we have recovered the moduli space of D3-branes in F-theory which indeed leads to the above geometry.

\begin{table}[!ht]
\begin{center}
\small
\begin{tabular}{|ccccccc|}\hline
 Name &  SW curve  & $\text{order}(\Delta)$ & $D(u)$ & Monodromy  & deficit angle & $\tau_0$   \\\hline\hline
$II^*$  & $y^2=x^3+u^5$ & $10$ & $6$ & $ST$ & $5\pi/3$ & $e^{i\pi/3}$     \\\hline
$III^*$  & $y^2=x^3+u^3x$ & $9$ & $4$ & $S$ & $3\pi/2$ & $i$     \\\hline
$IV^*$  &  $y^2=x^3+u^4$ & $8$ & $3$ & $-(ST)^{-1}$ & $4\pi/3$ & $e^{i\pi/3}$     \\\hline
$I_0^*$   &  $y^2=\prod_i^4(x-e_i(\tau)u)$ & $6$ & $2$ & $-I$ & $\pi$ & $\tau$    \\\hline
$IV$  & $y^2=x^3+u^2$ & $4$ & $3/2$ & $-ST$ & $2\pi/3$ & $e^{i\pi/3}$     \\\hline
$III$ & $y^2=x^3+ux$ & $3$ & $4/3$ & $S^{-1}$ & $\pi/2$ & $i$    \\\hline
$II$ & $y^2=x^3+u$ & $2$ & $6/5$ & $(ST)^{-1}$ & $\pi/3$ & $e^{i\pi/3}$  \\\hline\hline
$I_{n>0}^*$ & $y^2=x^3+ux^2+\Lambda^{-2n}u^{n+3}$ & $n+6$ & $2$ & $-T^n$ & cusp & $i\infty$     \\\hline
$I_{n>0}$  & $y^2=(x-1)(x^2+\Lambda^{-2n}u^n)$ & $n$ & $1$ & $T^n$ & cusp & $i\infty$     \\\hline
\end{tabular}
\caption{
The geometries around a singularity in the Coulomb branch are listed. This table is taken from \cite{Argyres:2015ffa}. 
The Seiberg-Witten curve around the singularity is written in the second column where the singularity is at $u=0$. 
The third column, $\text{order}(\Delta)$, is the order of zero of the discriminant.
The fourth column, $D(u)$, is the scaling dimension of $u$. 
The fifth column is the monodromy up to $SL(2,\mathbb{Z})$ conjugation, where $S$ and $T$ are given in \eqref{Eq:monodromy}.
The sixth column is the deficit angle in Coulomb branch geometry, and the last column is the value of $\tau$ at the singular point.
The last two rows correspond to the IR free theories while the others are SCFTs. 
}
\label{Tab:Classification}
\end{center}
\end{table}

Apriori there is no reason that the elliptic fibrations are smooth.  Indeed, as we will see in the next section the singularities are not only allowed, but are in fact required when there are non-abelian gauge symmetries.
The possible singularities of elliptic fibration are classified by Kodaira~\cite{10.2307/1970131,10.2307/1970500}. 
The corresponding geometry of the Coulomb branch in 4d $\mathcal{N}=2$ rank one theory is given in Table~\ref{Tab:Classification} which is taken from \cite{Argyres:2015ffa}.
There are nine types of singularities, $II^*, III^*, IV^*, I_0^*, IV, III, II, I_{n>0}^*$ and $I_{n>0}$.
Except for $I^*_{n>0}$ and $I_{n>0}$, the singular points correspond to SCFTs where a dimensional parameter does not appear in the Seiberg-Witten curve. 
These singular points are the tips of the cone, which gives rise to the deficit angle presented in Table~\ref{Tab:Classification}.
On the other hand, Seiberg-Witten curves of $I_{n>0}$ and $I^*_{n>0}$ singularities contain a dimensional parameter $\Lambda$, which indicates that these points are IR free theory rather than SCFT. The parameter $\Lambda$ is the Landau pole.
In terms of Coulomb branch geometry, $I^*_{n>0}$ and $I_{n>0}$ singularities are cusps.
By enclosing the singularities, $\tau$ receives monodromy denoted in Table~\ref{Tab:Classification}, where
\begin{align}\label{Eq:monodromy}
&S=\begin{pmatrix}
0 & -1 \\
1 & 0
\end{pmatrix},
&&T=\begin{pmatrix}
1 & 1 \\
0 & 1
\end{pmatrix}.
\end{align}

The singular points in Table~\ref{Tab:Classification} contribute to the integral of curvature in \eqref{Eq:inequality} for
 $II^*, III^*, IV^*, I_0^*$, $IV, III$, and $II$ singularities as delta-function sources leading to deficit angles noted in Table~\ref{Tab:Classification}.
On the other hand, $I^*_{n>0}$ and $I_{n>0}$ singularities contribute to \eqref{Eq:inequality} not as a delta-function but as an integral.
The profile of $\tau$ induced by the $I_{n>0}$ singularity at $u=0$ takes a form~\cite{Greene:1989ya}
\begin{align}
j(\tau)\sim \frac{1}{u^n},
\end{align}
where $j$ is the modular invariant function, 
Note that, around $u\sim0$, the behavior of $\tau$ is
\begin{align}
\tau \sim \frac{n}{2\pi i} \log u,
\end{align}
which reproduces the monodromy $T^n$ in Table~\ref{Tab:Classification}.
The contribution to \eqref{Eq:inequality} is computed as~\cite{Greene:1989ya}
\begin{align}\label{Eq:tau_integral}
-\frac{i}{2}\left. \int  \frac{\partial\tau \overline{\partial} \bar{\tau}}{(\tau - \bar{\tau})^2}\right|_{I_n}
= n\frac{\pi}{12}.
\end{align}
The contribution from $I^*_{n>0}$ singularity is computed in a similar way.
In addition to the contribution like \eqref{Eq:tau_integral}, this singularity generates the deficit angle $\pi$ reflecting the factor $(-1)$ in the monodromy matrix. 

Due to the positivity of the integral in \eqref{Eq:inequality} the contribution of the singular points should be less than or equal to $4\pi$.
Putting all together, we obtain the bound on the number/type of singular points as
\begin{align}\label{Eq:singularity_bound}
&10\,\#(II^*) + 9\,\#(III^*) + 8\,\#(IV^*) + 4\,\#(IV)
\nonumber \\& + 3\,\#(III)+ 2\, \#(II) +\sum_{n=0} \bigg(n \#(I_{n})+(n+6)\#(I^*_n)\bigg)\leq 24.
\end{align}

\subsection{The relation between the type of the singularity and the gauge algebra}
Here we provide the relation between the type of singularity and worldvolume flavor symmetry in order to write the bound \eqref{Eq:singularity_bound} in terms of the bulk gauge symmetry. 
In general, the type of singularity does not uniquely determine the flavor symmetry.
However, the requirement that the Higgs branch is the one-instanton moduli space is a strong input to obtain such a relation.

As we reviewed in Sec~\ref{Sec:swampland_review}, the gauge algebras allowed by the Swampland bounds so far are $\mathfrak{su}(n), \mathfrak{so}(2n), \mathfrak{sp}(n), \mathfrak{e}_{6,7,8}$, and $\mathfrak{g}_2$.
In the following, we first discuss the bound on the simply-laced gauge algebra. Then, $\mathfrak{sp}(n)$ and $\mathfrak{g}_2$ are discussed in Sections \ref{Sec:sp(n)} and \ref{Sec:g2}, respectively. 

In Table~\ref{Tab:Higgs_Branch}, the central charges and $D(u)$ of SCFTs with flavor symmetry $\mathfrak{g}$ are listed by assuming that the Higgs branch is the one-instanton moduli space. By comparing the value of $D(u)$ in Tables~\ref{Tab:Classification} and \ref{Tab:Higgs_Branch}, we obtain the relation between the flavor symmetry and the type of the singularity as follows.
\begin{align}\label{Eq:flavor_symmetry}
&II^* \leftrightarrow \mathfrak{e}_8,
&&III^* \leftrightarrow \mathfrak{e}_7,
&&IV^* \leftrightarrow \mathfrak{e}_6,
&&I_0^* \leftrightarrow \mathfrak{so}(8),
\nonumber\\
&IV \leftrightarrow \mathfrak{su}(3),
&&III \leftrightarrow \mathfrak{su}(2),
&&II \leftrightarrow \varnothing,
\end{align}
The SCFTs with $\mathfrak{e}_{6,7,8}$ are Minahan-Nemeschasky theories~\cite{Minahan:1996fg,Minahan:1996cj}, and the ones with $\mathfrak{su}(2)$ and $\mathfrak{su}(3)$ (and $\varnothing$) are Argyres-Douglas SCFT~\cite{Argyres:1995jj,Argyres:1995xn}.
As for the IR free theories (the last two rows in Table~\ref{Tab:Classification}), the $I_{n>0}$ singularity is interpreted as $\mathfrak{u}(1)$ gauge theory with $n$ electrons, and $I_{n>0}^*$ singularity is interpreted as $\mathfrak{su}(2)$ gauge theory with $n$ quarks:
\begin{align}\label{Eq:flavor_symmetry_2}
&I_{n>0}^* \leftrightarrow  \mathfrak{so}(2n),
&&I_{n>0} \leftrightarrow  \mathfrak{su}(n).
\end{align}

In terms of the simply-laced gauge algebra, we have
 \begin{align}\label{Eq:result}
 &\sum_{i} \bigg(a_i \,\mathfrak{su}(n_i)
 \oplus d_i \,\mathfrak{so}(2m_i)\bigg)
 \oplus l_1 \mathfrak{e}_6 \oplus l_2 \mathfrak{e}_7 \oplus l_3 \mathfrak{e}_8
\nonumber\\
&\text{with } \sum_i \bigg(n_i a_i + (m_i+2)d_i\bigg) + 8l_1 + 9l_2 + 10 l_3 \leq24,
 \end{align}
where $a_i, d_i, l_{1,2,3}$ are non-negative integers. 
Although we derive this without using string theory, the bound above has a clear interpretation in F-theory.
This is nothing but the statement that there are 24 sevenbranes~\cite{Vafa:1996xn, Morrison:1996na, Morrison:1996pp}.

Dividing both sides by $24$, we obtain the values of the last row in Table~\ref{Tab:Group}, where order $\Delta$ is given in Table~\ref{Tab:Classification} and (\ref{Eq:flavor_symmetry}, \ref{Eq:flavor_symmetry_2}).
By comparing the last three rows, we observe that the 3-brane bound is stronger for smaller gauge algebra, and the 1-brane bound is stronger for bigger gauge algebra.

\subsection{\texorpdfstring{$\mathfrak{sp}(n)$}{} gauge algebra}
\label{Sec:sp(n)}
As we have seen, the instanton 3-brane is a powerful object to obtain Swampland constraints.
However, we should be careful for $\mathfrak{sp}$ gauge algebra.  An $\mathfrak{sp}(n)$ gauge symmetry in the bulk would lead on the instanton brane to a theory with $\mathfrak{sp}(n)$ flavor symmetry and the one-instanton moduli space as the Higgs branch.
This is the theory with just one $\mathfrak{sp}(n)$ half-hypermultiplet.
This suffers from Witten's global gauge anomaly~\cite{Witten:1982fp} once we gauge it. 
This signals the bulk $\mathfrak{sp}(n)$ symmetry is anomalous.
This type of subtle global gauge anomaly is found in \cite{Garcia-Etxebarria:2017crf}, where it is argued that this anomaly is canceled by the topological Green-Schwarz mechanism. 
Namely, the 8d supergravity must be coupled with TQFT which gives the constraint that the $\mathfrak{sp}(n)$ instanton number is even~\cite{Pantev:2005rh,Pantev:2005wj,Pantev:2005zs,Hellerman:2006zs,Seiberg:2010qd,Tanizaki:2019rbk}.

Therefore, probe worldvolume theory with $\mathfrak{sp}(n)$ symmetry corresponding to a singular point is expected to have two-instantons moduli space as the Higgs branch.  This leads as noted before to
\begin{align}\label{Eq:level_2}
\ell_{ADE}=2\ell_{\mathfrak{sp}}.
\end{align}
The case of $\mathfrak{su}(2)=\mathfrak{sp}(1)$ is subtle. If the TQFT enforces the instanton number to be even, we should take $\ell=\ell_{\mathfrak{sp}}$ otherwise $\ell=\ell_{ADE}$.

The detail of the worldvolume theory is discussed in F-theory case~\cite{Witten:1997bs}, see also \cite{Argyres:2016yzz} for a study in the context of field theory.
When $D3$-brane hits the frozen $O7_+$ plane, the worldvolume theory is $O(2)$ gauge theory with two massless charge $2$ hypermultiplets. 
The monodromy is $-T^4$ up to $SL(2,\mathbb{Z})$ conjugation. The $(-1)$ factor arises from $\mathbb{Z}_2$ gauging and $T^4$ is understood as the beta function coming from hypermultiplets. 
Therefore, this is the $I_4^*$ singularity.
Moreover, the mass deformation is not possible due to $\mathbb{Z}_2$ gauging, and this is a frozen $I_4^*$ singularity.
Putting $n$ $D7$-brane on top of $O7_+$ plane, $\mathfrak{sp}(n)$ gauge algebra is realized.
The level of $\mathfrak{sp}(n)$ is one while the level of simply-laced gauge algebra is two.
The $D3$-brane can fractionate into two separate objects on $O7_+$ plane~\cite{Bhardwaj:2018jgp}, corresponding to the two-instantons moduli space Higgs branch.

With this interpretation, we obtain the bound including $\mathfrak{sp}$ as\footnote{We assume no other theories with the two-instantons moduli space since there are no other examples in the classification of 4d $\mathcal{N}=2$ rank one theories~\cite{Argyres:2015ffa,Argyres:2015gha,Argyres:2016xua,Argyres:2016xmc,Argyres:2016yzz}}
 \begin{align}\label{Eq:result_2}
 &\sum_{i} \bigg(a_i \,\mathfrak{su}(n_i)
 \oplus d_i \,\mathfrak{so}(2m_i) 
 \oplus f_i \, \mathfrak{sp}(k_i) \bigg)
 \oplus l_1 \mathfrak{e}_6 \oplus l_2 \mathfrak{e}_7 \oplus l_3 \mathfrak{e}_8
\nonumber\\
&\text{with } \sum_i \bigg(n_i a_i + (m_i+2)d_i + (10+k_i)f_i\bigg) + 8l_1 + 9l_2 + 10 l_3 \leq24,
 \end{align}
where the $\mathfrak{sp}(n)$ is related to the type of singularity as
\begin{align}
I_{4+n}^* \leftrightarrow  \mathfrak{sp}(n),
\end{align}

\subsection{\texorpdfstring{$\mathfrak{g}_2$}{} gauge theory}
\label{Sec:g2}

Finally, we discuss $\mathfrak{g}_2$ gauge algebra.
As discussed in Section~\ref{Sec:swampland_review}, $\mathfrak{g}_2$ may appear only in Rank $=10$ theories (and possibly Rank $=2$ theories with $\kappa=1$).
We argue that $\mathfrak{g}_2$ does not appear in these cases either.\footnote{We thank Mario Martone for sharing this observation with us.}
In other words supergravity theories with $\mathfrak{g}_2$ symmetry in 8d are in the Swampland.

There is a relation~\cite{Shapere:2008zf,Argyres:2016xmc}
 \begin{align}\label{Eq:relation}
4(2a-c)=\sum_{i=1}^r2 D(u_i)-r
 \end{align}
where $r$ is the rank of the theory. 
By combining the unitarity bound~\cite{Mack:1975je,Dobrev:1985qv}\footnote{It is assumed here that there are no ``complex singularities''~\cite{Argyres:2017tmj}. There are no known examples of SCFTs violating this assumption.} $D(u_i)\geq1$ with $a=17/24, c=5/6$ in Table~\ref{Tab:Higgs_Branch}, the value of $r$ is $ 1$ or $2$.
For rank-$1$ case, we obtain $D(u)=5/3$, but this value does not appear in Table~\ref{Tab:Group}.
For rank-2, we obtain $D(u_1)+D(u_2)=13/6$. 
The allowed scaling dimensions for scale invariant geometries of rank-2 are known \cite{Caorsi:2018zsq,Argyres:2018urp}. 
It turns out that there are no pairs which add up to $13/6$.


In the above discussion, once we assume $\mathcal{N}=2$ supersymmetry on the 3-brane worldvolume as required by the BPS completeness hypothesis, we only used local features of 4d $\mathcal{N}=2$ theories to rule out $\mathfrak{g}_2$ gauge theory with 16 supercharges in 8d regardless of coupling to gravity. 

\section{Power of the geometry and the bound}
\label{Sec:power}

The geometry of the instanton 3-brane is powerful in constraining the allowed gauge groups.  For rank $18$, since the geometry of the 3-brane moduli space is identical to that of F-theory and these are arbitrary elliptic K3 geometries, we obtain exactly the same gauge groups that are allowed in the string constructions.  In the ranks 10 and 2 cases, the instanton 3-brane geometry and F-theory geometry both are given by elliptic K3 geometries.  However, the restrictions that the F-theory places on the allowed K3 geometries appearing are not manifest from the viewpoint of the instanton 3-branes.  Thus we cannot yet
obtain the exact same restrictions.  It is natural to expect that some refinement of the discussion here, would lead to this full restriction of the K3 geometries appearing in F-theory for these cases as well and reproduce the exact match with string constructions.  

In the absence of this, we can still use
 the bounds (\ref{Eq:result}, \ref{Eq:result_2}) to constrain the actual gauge algebra in the lower rank theories as well.
 In order to illustrate the power of the bound, it is interesting to compare the number of models allowed by the Swampland constraint and string constructions. 
Even though as we have argued in the rank 18 cases we reproduce exactly the observed gauge groups in string theory using the knowledge of the geometry, it is interesting to see how much of this match can be explained using the simpler restriction on the bound.
We consider this question first, before turning to the rank 10 and 2 cases.

First, the patterns of the gauge algebra we know in string theory is
\begin{align}
&\text{\#(String construction)}=3279,
\end{align}
except for the trivial $18\,\mathfrak{u}(1)$.
On the other hand, from the Swampland bounds reviewed in Section~\ref{Sec:swampland_review}, only simply-laced algebra is allowed for the rank $=18$ theories from the unitarity of the $1$-brane probe. 
The number of simply-laced gauge algebra whose total rank is equal or less than $18$ is
\begin{align}
&\text{\#(Consistent with the unitarity of $1$-brane probes)}=5366.
\end{align}
Therefore, there are $5366-3279=2087$ patterns that are consistent with the unitarity of $1$-brane probes, but without known string construction. 
From \eqref{Eq:result}, $3$-brane condition says that some of them are in the Swampland:
\begin{align}
&\text{\#(Excluded by $3$-brane probes)}=1429.
\end{align}

When we focus on the models without $\mathfrak{u}(1)$, we obtain the following numbers.
\begin{align}
&\text{\#(String construction)}=325,
\nonumber\\
&\text{\#(Consistent with the unitarity of $1$-brane probes)}=1599,
\nonumber\\
&\text{\#(Excluded by $3$-brane probes)}=887,
\nonumber\\
&\text{\#(Consistent with $1$-brane and $3$-brane probes)}=712.
\end{align}

\begin{table}[!ht]
\begin{center}
\small
\begin{tabular}{|c||ccc|}\hline
 $\mathfrak{g}$ &  \#(string construction)  & \#(rank $10$ patterns)  & order$(\Delta)/24$ \\\hline\hline
$\mathfrak{sp}(2)\oplus A$  & $1$ & $1$ & $21/24$ \\\hline
$\mathfrak{sp}(2)\oplus 2A$  & $3$ & $4$ & $22/24$  \\\hline
$\mathfrak{sp}(2)\oplus 3A$   & $1$  & $5$ &  $23/24$ \\\hline
$\mathfrak{sp}(2)\oplus 4A$   & $2$  & $5$ &  $24/24$ \\\hline
$\mathfrak{sp}(2)\oplus 5A$  & $0$ & $3$ &  $25/24$ \\\hline
$\mathfrak{sp}(2)\oplus 6A$  &  $0$ & $2$ & $26/24$  \\\hline
$\mathfrak{sp}(2)\oplus 7A$  & $0$ & $1$ & $27/24$  \\\hline
$\mathfrak{sp}(2)\oplus 8A$  & $0$ & $1$ & $28/24$  \\\hline
total  & $7$ & $22$ & $7$ models are excluded  \\\hline
\end{tabular}
\caption{
As an illustration of the power of $3$-brane bound, we show a subset of the gauge algebras $\mathfrak{sp}(2)\oplus nA$, where the total rank is $10$ and no $\mathfrak{u}(1)$ factor. 
The second column is the number of patterns realized by F-theory on elliptic $K3$ with one frozen singularity. 
The third column is the number of patterns consistent with the $1$-brane probe bound~\cite{Kim:2019ths}. The last column shows the patterns excluded by the $3$-brane probe bound \eqref{Eq:result}, order$(\Delta)/24\leq1$. 
}
\label{Tab:K3}
\end{center}
\end{table}

\begin{table}[!ht]
\begin{center}
\footnotesize
\begin{tabular}{|c||ccc|}\hline
 $\mathfrak{g}$ &      singularity &  \#($7$-brane)$_{ADE}$ & No. in \cite{2005math......5140S} \\\hline\hline
$\mathfrak{e}_8 \oplus\mathfrak{sp}(2)$  &  $I_6^*+II^*$ & $22$ & $2961$ \\\hline
$\mathfrak{e}_8\oplus\mathfrak{sp}(1)\oplus\mathfrak{su}(2)$  &  $(III\text{ or } I_2)+I_5^*+II^*$ & $\geq22$ & $2962$\\\hline
$\mathfrak{e}_7\oplus\mathfrak{sp}(3)$  &  $I_7^*+III^*$ & $22$ & $2992$ \\\hline
$\mathfrak{e}_7\oplus\mathfrak{sp}(2)\oplus\mathfrak{su}(2)$  &   $I_2+I_6^*+III^*$ & $23$ & $2993$\\\hline
$\mathfrak{e}_7\oplus\mathfrak{sp}(1)\oplus\mathfrak{su}(3)$  &   $I_3 + I_5^* + III^*$ & $23$ & $2994$\\\hline
$\mathfrak{e}_7\oplus\mathfrak{su}(3)\oplus\mathfrak{su}(2)$  &   $I_2+I_3+III^*$ & $24$ & $2995$\\\hline
$\mathfrak{e}_6\oplus\mathfrak{sp}(4)$  &   $I_8^*+IV^*$ & $22$ & $3029$\\\hline
$\mathfrak{e}_6\oplus\mathfrak{sp}(3)\oplus\mathfrak{su}(2)$  & $I_2+I_7^*+IV^*$ & $23$ & $3030$ \\\hline
$\mathfrak{e}_6\oplus\mathfrak{sp}(1)\oplus\mathfrak{su}(4)$  & $I_4+I_5^*+IV^*$ & $23$ & $3031$ \\\hline
$\mathfrak{e}_6\oplus\mathfrak{sp}(1)\oplus\mathfrak{su}(3)\oplus\mathfrak{su}(2)$  & $I_2+I_3+I_5^*+IV^*$ & $24$ & $3032$ \\\hline
$\mathfrak{e}_6\oplus\mathfrak{su}(5)$   & $I_5+I_4^*+IV^*$ & $23$ & $3033$\\\hline
$\mathfrak{sp}(10)$   & $I_{14}^*$ & $20$ & $3061$ \\\hline
$\mathfrak{sp}(9)\oplus\mathfrak{su}(2)$ & $(III\text{ or } I_2)+I_{13}^*$ & $\geq21$ & $3062$ \\\hline
$\mathfrak{sp}(8)\oplus\mathfrak{su}(3)$ & $(IV\text{ or } I_3)+I_{12}^*$ & $\geq21$ & $3063$ \\\hline
$\mathfrak{sp}(8)\oplus2\,\mathfrak{su}(2)$ & $2(III\text{ or } I_2)+I_{12}^*$ & $\geq22$ & $3064$\\\hline
$\mathfrak{sp}(7)\oplus\mathfrak{su}(3)\oplus\mathfrak{su}(2)$ & $(III\text{ or } I_2)+(IV\text{ or } I_3)+I_{11}^*$ & $\geq22$ & $3065$ \\\hline
$\mathfrak{sp}(6)\oplus\mathfrak{su}(5)$ & $I_5+I_{10}^*$ & $21$ & $3066$ \\\hline
$\mathfrak{sp}(6)\oplus\mathfrak{su}(4)\oplus\mathfrak{su}(2)$ & $(III\text{ or } I_2)+I_4+I_{10}^*$ & $\geq22$ & $3067$ \\\hline
$\mathfrak{sp}(6)\oplus2\,\mathfrak{su}(3)$ & $2(IV\text{ or } I_3)+I_{10}^*$ & $\geq22$ & $3068$\\\hline
$\mathfrak{sp}(6)\oplus\mathfrak{su}(3)\oplus2\,\mathfrak{su}(2)$ & $2I_2+I_3+I_{10}^*$ & $23$ & $3069$ \\\hline
$\mathfrak{sp}(5)\oplus\mathfrak{so}(10)$ & $I_1^*+I_9^*$ & $22$ & $3070$ \\\hline
$\mathfrak{sp}(5)\oplus\mathfrak{su}(6)$ & $I_6+I_9^*$ & $21$ & $3071$\\\hline
$\mathfrak{sp}(5)\oplus\mathfrak{su}(5)\oplus\mathfrak{su}(2)$ & $(III\text{ or } I_2)+I_5+I_9^*$ & $\geq22$ & $3072$\\\hline
$\mathfrak{sp}(4)\oplus\mathfrak{so}(12)$ & $I_2^*+I_8^*$ & $22$ & $3073$\\\hline
$\mathfrak{sp}(4)\oplus\mathfrak{so}(10)\oplus\mathfrak{su}(2)$ & $I_2+I_1^*+I_8^*$ & $23$ & $3074$\\\hline
$\mathfrak{sp}(4)\oplus\mathfrak{su}(5)\oplus2\,\mathfrak{su}(2)$ & $2I_2+I_5+I_8^*$ & $23$ & $3075$\\\hline
$\mathfrak{sp}(4)\oplus\mathfrak{su}(4)\oplus\mathfrak{su}(3)\oplus\mathfrak{su}(2)$ & $I_2+I_3+I_4+I_8^*$ & $23$ & $3076$ \\\hline
$\mathfrak{sp}(4)\oplus2\,\mathfrak{su}(3)\oplus2\,\mathfrak{su}(2)$ & $2I_2+2I_3+I_8^*$ & $24$ & $3077$ \\\hline
$\mathfrak{sp}(3)\oplus\mathfrak{su}(7)\oplus\mathfrak{su}(2)$ & $(III\text{ or } I_2)+I_7+I_7^*$ & $\geq22$ & $3078$ \\\hline
$\mathfrak{sp}(3)\oplus\mathfrak{su}(6)\oplus\mathfrak{su}(3)$ & $(IV\text{ or } I_3)+I_6+I_7^*$ & $\geq22$ & $3079$ \\\hline
$\mathfrak{sp}(3)\oplus\mathfrak{su}(5)\oplus\mathfrak{su}(3)\oplus\mathfrak{su}(2)$ & $I_2+I_3+I_5+I_7^*$ & $23$ & $3080$ \\\hline
$\mathfrak{sp}(3)\oplus\mathfrak{su}(4)\oplus2\,\mathfrak{su}(3)$  & $2I_3+I_4+I_7^*$ & $23$ & $3081$ \\\hline
\end{tabular}
\caption{
List of Rank $=10$ theories realized in string theory (Part $1/2$). 
According to \cite{Witten:1997bs}, $O7_+$ plane plus $n$ $7$-brane is described as the frozen $I_{n+4}^*$ singularity in F-theory, and gives rise to $\mathfrak{sp}(n)$ symmetry.
By using a list of singularities in \cite{2005math......5140S}, we obtain Tables~\ref{Tab:Rank10}, \ref{Tab:Rank10_2} and \ref{Tab:Rank2}. The correspondence with Table 1 in \cite{2005math......5140S} is shown in last column.
} 
\label{Tab:Rank10}
\end{center}
\end{table}
\begin{table}[!ht]
\begin{center}
\footnotesize
\begin{tabular}{|c||ccc|}\hline
 $\mathfrak{g}$ &    singularity &  \#($7$-brane)$_{ADE}$ & No. in \cite{2005math......5140S} \\\hline\hline
 $\mathfrak{sp}(2)\oplus\mathfrak{so}(14)\oplus\mathfrak{su}(2)$ & $I_2+I_3^*+I_6^*$ & $23$ & $3082$ \\\hline
$\mathfrak{sp}(2)\oplus\mathfrak{so}(12)\oplus\mathfrak{su}(3)$ & $I_3+I_2^*+I_6^*$ & $23$ & $3083$ \\\hline
$\mathfrak{sp}(2)\oplus\mathfrak{so}(10)\oplus\mathfrak{su}(3)\oplus\mathfrak{su}(2)$ & $I_2+I_3+I_1^*+I_6^*$ & $24$ & $3084$ \\\hline
$\mathfrak{sp}(2)\oplus\mathfrak{su}(9)$ & $I_9+I_6^*$ & $21$ & $3085$\\\hline
$\mathfrak{sp}(2)\oplus\mathfrak{su}(7)\oplus\mathfrak{su}(3)$ & $(IV\text{ or } I_3)+I_7+I_6^*$ & $\geq22$ & $3086$\\\hline
$\mathfrak{sp}(2)\oplus\mathfrak{su}(6)\oplus\mathfrak{su}(4)$ & $I_4+I_6+I_6^*$ & $22$ & $3087$ \\\hline
$\mathfrak{sp}(2)\oplus\mathfrak{su}(6)\oplus3\,\mathfrak{su}(2)$ & $3I_2+I_6+I_6^*$ & $24$ & $3088$ \\\hline
$\mathfrak{sp}(2)\oplus2\,\mathfrak{su}(5)$ & $2I_5+I_6^*$ & $22$ & $3089$ \\\hline
$\mathfrak{sp}(2)\oplus\mathfrak{su}(5)\oplus\mathfrak{su}(4)\oplus\mathfrak{su}(2)$ & $I_2+I_4+I_5+I_6^*$ & $23$ & $3090$ \\\hline
$\mathfrak{sp}(2)\oplus2\,\mathfrak{su}(4)\oplus2\,\mathfrak{su}(2)$ & $2I_2+2I_4+I_6^*$ & $24$ & $3091$\\\hline
$\mathfrak{sp}(1)\oplus\mathfrak{so}(18)$ & $2I_5^*$ & $22$ & $3092$ \\\hline
$\mathfrak{sp}(1)\oplus\mathfrak{so}(10)\oplus\mathfrak{su}(5)$ & $I_5+I_1^*+I_5^*$ & $23$ & $3093$ \\\hline
$\mathfrak{sp}(1)\oplus\mathfrak{su}(10)$ & $I_{10}+I_5^*$ & $21$ & $3094$ \\\hline
$\mathfrak{sp}(1)\oplus\mathfrak{su}(9)\oplus\mathfrak{su}(2)$ & $(III\text{ or } I_2)+I_9+I_5^*$ & $\geq22$ & $3095$\\\hline
$\mathfrak{sp}(1)\oplus\mathfrak{su}(8)\oplus2\,\mathfrak{su}(2)$ & $2I_2+I_8+I_5^*$ & $23$ & $3096$\\\hline
$\mathfrak{sp}(1)\oplus\mathfrak{su}(7)\oplus\mathfrak{su}(3)\oplus\mathfrak{su}(2)$ & $I_2 + I_3+I_7+I_5^*$ & $23$ & $3097$ \\\hline
$\mathfrak{sp}(1)\oplus\mathfrak{su}(6)\oplus\mathfrak{su}(5)$ & $I_5+I_6+I_5^*$ & $22$ & $3098$\\\hline
$\mathfrak{sp}(1)\oplus\mathfrak{su}(6)\oplus\mathfrak{su}(4)\oplus\mathfrak{su}(2)$ & $I_2+I_4+I_6+I_5^*$ & $23$ & $3099$ \\\hline
$\mathfrak{sp}(1)\oplus\mathfrak{su}(5)\oplus2\,\mathfrak{su}(3)\oplus\mathfrak{su}(2)$ & $I_2+2 I_3+I_5+I_5^*$ & $24$ & $3100$\\\hline
$\mathfrak{so}(16)\oplus2\,\mathfrak{su}(2)$ & $2I_2+2I_4^*$ & $24$ & $3101$ \\\hline
$\mathfrak{so}(12)\oplus\mathfrak{su}(4)\oplus\mathfrak{su}(2)$ & $I_2+I_4+2I_4^*$ & $24$ & $3102$\\\hline
$2\,\mathfrak{so}(10)$ & $2I_1^*+I_4^*$ & $24$ & $3103$ \\\hline
$\mathfrak{su}(10)\oplus\mathfrak{su}(2)$ & $(III\text{ or } I_2)+I_{10}+I_4^*$ & $\geq21$ & $3104$\\\hline
$\mathfrak{su}(8)\oplus\mathfrak{su}(3)\oplus\mathfrak{su}(2)$ & $I_2+I_3+I_8+I_4^*$ & $23$ & $3105$\\\hline
$\mathfrak{su}(7)\oplus2\,\mathfrak{su}(3)$ & $2I_3+I_7+I_4^*$ & $23$ & $3106$\\\hline
$2\,\mathfrak{su}(6)$ & $2I_6+I_4^*$ & $22$ & $3107$\\\hline
$\mathfrak{su}(6)\oplus\mathfrak{su}(5)\oplus\mathfrak{su}(2)$ & $I_2+I_5+I_6+I_4^*$ & $23$ & $3108$\\\hline
$\mathfrak{su}(6)\oplus\mathfrak{su}(4)\oplus2\,\mathfrak{su}(2)$ & $2I_2+I_4+I_6+I_4^*$ & $24$ & $3109$\\\hline
$2\,\mathfrak{su}(4)\oplus2\,\mathfrak{su}(3)$ & $2I_3+2I_4+I_4^*$ & $24$ & $3110$\\\hline
\end{tabular}
\caption{
List of Rank $=10$ theories realized in string theory (Part $2/2$).  
}
\label{Tab:Rank10_2}
\end{center}
\end{table}

For the rank 10, the patterns of the gauge algebra in string theory are (see Tables~\ref{Tab:Rank10} and \ref{Tab:Rank10_2} for the gauge algebras of maximal enhancement)\footnote{We thank the authors of \cite{Font:2021uyw} for discussion.}
\begin{align}
&\text{\#(rank 10 string construction)}=327
\end{align}
except for $10\,\mathfrak{u}(1)$.
Here we do not distinguish the model having different levels.
The simply-laced, $\mathfrak{sp}(n)$ and $\mathfrak{g}_2$ gauge algebra are allowed as long as \eqref{Eq:1-brane} is satisfied.
The number of patterns whose total rank is equal or less than $10$ is
\begin{align}
&\text{\#(Consistent with the unitarity of $1$-brane probes)}=997.
\end{align}
The bound~\eqref{Eq:result_2} and argument in Section~\ref{Sec:g2} gives
\begin{align}
&\text{\#(Excluded by $3$-brane probes)}=522,
\nonumber\\
&\text{\#(Consistent with $1$-brane and $3$-brane probes)}=475.
\end{align}
In table~\ref{Tab:K3}, we show the detailed comparison among \#(String construction), \#(Consistent with the unitarity of $1$-brane probes), and \#(Excluded by $3$-brane probes), concentrating on the subset of the algebra.

\begin{table}[!ht]
\begin{center}
\small
\begin{tabular}{|c||cccc|}\hline
 $\mathfrak{g}$ &  $c_\mathfrak{g}$  &  singularity &  \#($7$-brane)$_{ADE}$  & No. in \cite{2005math......5140S} \\\hline\hline
$2\,\mathfrak{u}(1)$  & $2$ & $2I_4^*$ & $20$ & $1958$ \\\hline
$\mathfrak{sp}(1)_1\oplus \mathfrak{u}(1)$  & $2$ & $I_4^* + I_5^*$ & $21$ & $2584$ \\\hline
$\mathfrak{su}(2)_1 \oplus \mathfrak{u}(1)$  & $2$ & $(III\text{ or } I_2) + 2I_4^*$ & $\geq22$ & $2604$ \\\hline
$2\mathfrak{sp}(1)_1$  & $2$ & $2 I_5^*$ & $21$ & $3092$ \\\hline
$2\mathfrak{su}(2)_1$  & $2$ & $2I_2 +2I_4^*$ & $24$ & $3101$ \\\hline
\end{tabular}
\caption{
The gauge algebras of theories with Rank $=2$. 
The subscript is the level of the algebra, which is determined by the $1$-brane bound $c_\mathfrak{g}\leq2$.
Note that $\mathfrak{sp}(1)$ comes from $I_5^*$ singularity, and $\mathfrak{su}(1)$ comes from $I_2$ singularity.
}
\label{Tab:Rank2}
\end{center}
\end{table}
For the rank 2 case with $\kappa=0$, the patterns of the gauge algebra in string theory are (see  Table~\ref{Tab:Rank2})
\begin{align}
&\mathfrak{su}(2), 
&&2\,\mathfrak{su}(2),
\end{align}
On the other hand, the patterns of simply-laced gauge algebra whose total rank is equal or less than $2$ are
\begin{align}
&\mathfrak{su}(2), 
&&2\,\mathfrak{su}(2),
&&\mathfrak{su}(3).
\end{align}
All the three patterns are consistent with the bound~\eqref{Eq:result_2}. 

\section{Conclusions}
\label{Sec:conclusions}
In this paper, we have investigated the Swampland bound of the 8d supergravity with 16 supercharges.
The key object of our discussion is the study of the instanton 3-brane moduli.
Furthermore, from the stronger form of the cobordism conjecture, we have argued that all the one-instanton configurations must be connected via a one dimensional Coulomb branch on the 3-brane probe.
We have derived the bound on the gauge algebra from the consistency conditions of the Coulomb branch geometry of the instanton 3-brane probes, which as we have seen leads to the reconstruction of the internal string geometry given by elliptic K3.
We have also shown why theories with $\mathfrak{g}_2$ symmetry are excluded.

In this paper, we have concentrated on gauge algebra, rather than the gauge group. 
Namely, we have not discussed the global structure of the gauge group. 
It is interesting to consider if we obtain the bound on the global structure of the gauge group from instanton 3-brane.
(See \cite{Cvetic:2020kuw,Montero:2020icj,Dierigl:2020lai} for discussions of the global structure of gauge groups).

The instanton $(d-4)$-branes are interesting objects to obtain new Swampland bounds.
It is interesting to investigate the instanton $4$-brane in 9d, $2$-brane in 7d, and so on.

As for other applications of the ideas considered in this paper, we expect studying brane probes of various types for different effective theories, can lead to reconstruction of the internal geometry of the would be string constructions.  This can open up a more efficient method to prove the SLP principle.
Also, the compactness of the brane probe moduli which we argued based on finiteness of BH entropy may have other extensions including to the distance conjecture which would be natural to study further.

\subsection*{Acknowledgments}
We would like to thank Philip Argyres, H\'ector Parra De Freitas, Bernardo Fraiman, Mariana Gra\~na, Mario Martone, Jacob McNamara and Miguel Montero for valuable discussions.

The work of YH is supported by JSPS Overseas Research Fellowships and the work of CV was partly supported by the
National Science Foundation under Grant No. NSF PHY-2013858.

\bibliographystyle{TitleAndArxiv}
\bibliography{Bibliography}

\end{document}